\documentclass[twocolumn,showpacs,preprintnumbers,amsmath,amssymb]{revtex4}

\usepackage{epsfig}
\usepackage{graphicx}
\usepackage{dcolumn}
\usepackage{bm}

\def\DESepsf(#1 width #2){\epsfxsize=#2 \epsfbox{#1}}
\newcommand{\im}{{\rm Im}}

\begin{document}


\title{Impact of Subleading Corrections on Hadronic $B$ Decays}

\author{Kwei-Chou Yang}

 \affiliation{Department of Physics, Chung
Yuan Christian University, Chung-Li, Taiwan 320, Republic of China
}%

\date{\today}

\begin{abstract}
We study the subleading corrections originating from the 3-parton
($q\bar q g$) Fock states of final-state mesons in $B$ decays. The
corrections could give significant contributions to decays
involving an $\omega$ or $\eta^{(\prime)}$ in the final states.
Our results indicate the similarity of $\omega K$ and $\omega
\pi^-$ rates, of order $5\times 10^{-6}$, consistent with the
recent measurements. We obtain $a_2(B\to J/\psi K)\approx
0.27+0.05i$, in good agreement with data. Without resorting to the
unknown singlet annihilation effects, 3-parton Fock state
contributions can enhance the branching ratios of $K\eta'$ to the
level above $ 50\times 10^{-6}$.
\end{abstract}

\maketitle


The rare B decays allow us to access the Kobayashi-Maskawa (KM)
mixing angles and search for new physics. Much progress in the
study of B decays~\cite{Beneke:2001ev,Keum:2000ph,Beneke:2002nj}
has been recently made in QCD-based approaches. In the
perturbative QCD (pQCD) framework, the importance of the weak
annihilation effects in $B\to K\pi$ decays was first emphasized by
\cite{Keum:2000ph}, where the annihilation contributions are
almost pure imaginary and therefore could lead to CP asymmetry
predictions different from the QCD factorization (QCDF)
results~\cite{Beneke:2001ev}. Nevertheless, the QCDF study showed
that the annihilation effects may play only a minor role in the
enhancement of $\pi\pi, \pi K$ branching ratios
(BRs)~\cite{Beneke:2001ev}. A recent QCDF fit to $K\pi, \pi\pi$
rates~\cite{Beneke:2002nj} indicated that even if the annihilation
contribution is neglected, one can still get quite good fitting
results provided that the strange quark mass is of order 80~MeV.

The annihilation effects might be much more important for $VP$
modes, where $P$ and $V$ denote pseudoscalar and vector mesons,
respectively.  It has been pointed out that in the absence of
annihilation effects, the $\phi K$ BRs are $\approx 4\times
10^{-6}$~\cite{Cheng:2000hv} which is too small compared to the
data $\sim 8\times 10^{-6}$~\cite{belle,babar_omegaK}. Recently
Belle observed a large $\omega K^-$ rate, $(6.7^{+1.3}_{-1.2}\pm
0.6)\times 10^{-6}$, and $\omega K^-/ \omega\pi^- \sim
1$~\cite{belle}. Sizable $\omega K$ results are also reported in
new BaBar measurements~\cite{babar_omegaK} with $\omega K^{-,0}
\sim \omega\pi^- \sim 5\times 10^{-6}$. It is hard to understand
the large strength of $\omega K$ rates from the theoretical point
of view. The ratio $\omega \overline K^0 / \omega \pi^-$ reads
\begin{eqnarray}
&& \omega \overline K^0/ \omega \pi^- \approx \left\vert V_{cb} /
V_{ub}\right\vert^2
\left(f_K / f_\pi \right)^2\nonumber\\
&&\times\left\vert \frac{a_4-a_6 r_\chi^K+ 2 r_2(a_3+a_5+a_9/4)
+f_B f_K b_3} { a_1+r_1 a_2} \right\vert^2, \hskip0.4cm {}
\end{eqnarray}
where $r_1 = f_\omega F_1^{B\pi} /f_{\pi} A_0^{B\omega}$,
$r_2=(F^{BK}_1 f_\omega)/(A^{B\omega}_0 f_K)$, $r^K_\chi=
2m_K^2/[m_b(m_s+m_u)]$ is the chirally enhanced factor with
$m_{s,u}$ being the current quark masses, and $b_3 \equiv
b_3(K,\omega)$ is the annihilation contribution defined
in~\cite{Du:2002up}. The $\omega \pi^-$ rate depends weakly on the
annihilation effects. Without annihilation, since $a_4$ and $a_6
r_\chi^K$ terms  in the $\omega \overline K^0 $ amplitude have
opposite signs, the ratio $\omega \overline K^0 /\omega \pi^-$
should be very small.  A possibility to explain the data is that
the annihilation effects may give the dominant contribution to
$\omega K$ modes as shown in the QCDF fit~\cite{Du:2002cf} for
$B\to PP$ together with some $B\to VP$ modes. (However, including
the contributions from annihilation effects, the pQCD results read
Br$(\overline B^0\to \omega \overline K^0)\lesssim 2\times
10^{-6}$~\cite{Chen:2001jx}.) This result hints that, to account
for the large $\omega \overline K^0$ rate, the annihilation
contributions to the BRs of all $B\to K V$ modes should be over
80\%. If it will be true, it should be easy to observe, for
instance, the following simple relation $\rho^+ K^{-,0}: \rho^0
K^{-,0}: \omega K^{-,0}\approx 1:(1/\sqrt{2})^2:(1/\sqrt{2})^2$,
the same as their annihilation ratios squared. Nevertheless, if
the global fit is extended to all measured $B\to PV$ modes, a
small $K\omega$ rate $\sim 2\times 10^{-6}$ will be
obtained~\cite{Aleksan:2003qi} and a reliable best fit cannot be
reached. The present QCD approach seems hard to offer a coherent
picture in dealing with $B\to VP$ modes.

\begin{figure}
\centerline{
        {\epsfxsize1.5in \epsffile{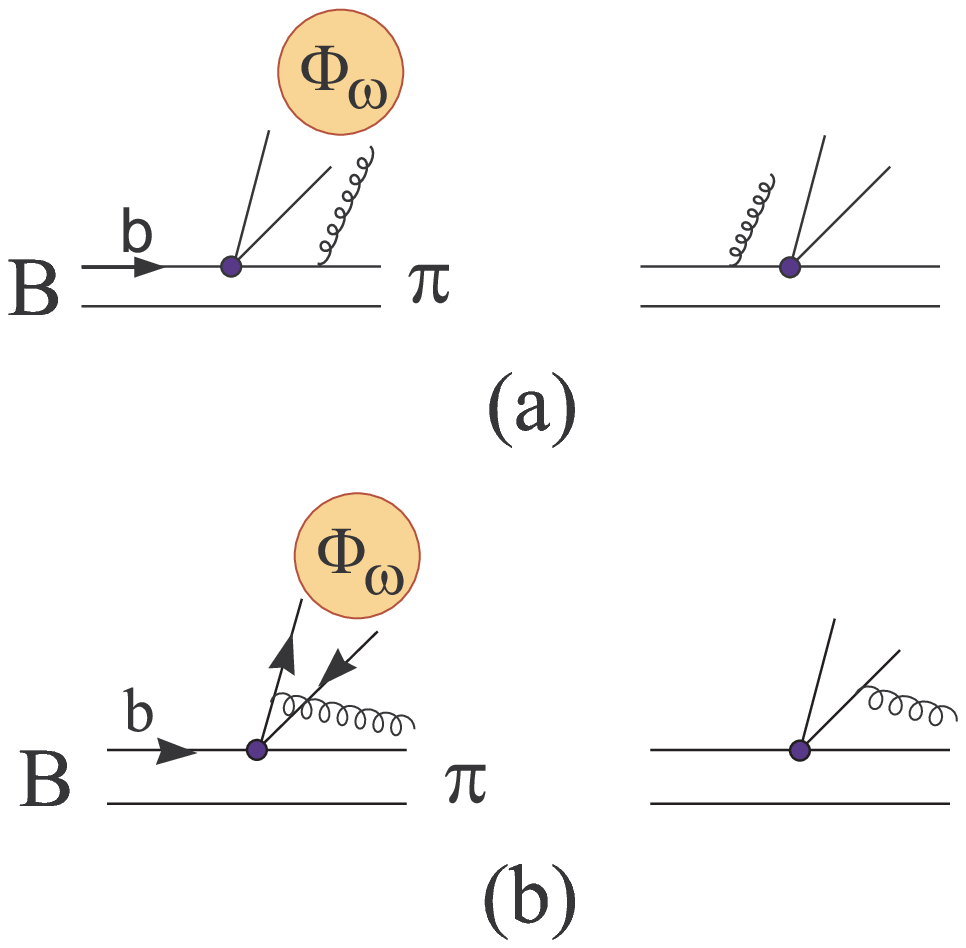}}}
\caption{The contributions of the $q \bar q g$ Fock states of the
(a) $\omega$ and (b) $\pi^-$ mesons to the $B^-\to \omega \pi^-$
amplitude.}\label{fig:3p}
\end{figure}

In this letter we take into account the subleading corrections
arising from the 3-parton Fock states of final state mesons, as
depicted in Fig.~\ref{fig:3p}, to QCDF decays amplitudes. We find
that it could give significant corrections to decays with
$\omega$, or $\eta^{(\prime)}$ in the final states. A simple rule
extended to $B\to PP,VP$ modes is obtained for the effective
coefficients $a_i^{\rm SL}$ with the subleading corrections,
 \begin{eqnarray}
 a^{\rm SL}_{2i}&=&a_{2i}+[1+(-1)^{\delta_{3i}+\delta_{4i}}]
 c_{2i-1}\, f_3/2,\nonumber\\
 a^{\rm SL}_{2i-1}&=&a_{2i-1}+ (-1)^{\delta_{3i}+\delta_{4i}}
 c_{2i}\, f_3,  \label{eq:wc}
 \end{eqnarray}
where $i=1,\cdots,5$, and $c_i$ are the Wilson coefficients
defined at the scale $\mu_h= \sqrt{\Lambda_\chi
m_B/2}\simeq1.4$~GeV with $\Lambda_\chi$ the momentum of the
emitted gluon as shown in Fig.~\ref{fig:3p}(b) and
\begin{eqnarray}
f_3&=& \frac{\sqrt{2}}{m_B^2 f_\omega F_1^{B\pi
}(m_\omega^2)}\langle \omega \pi^-|O_1|B^-\rangle_{\rm qqg} =0.12
\label{eq:p3}
\end{eqnarray}
in the SU(3) limit. Here $O_1=\overline s \gamma^\mu (1-\gamma_5)u
\, \overline u \gamma_\mu (1-\gamma_5) b$, and $\overline\alpha_g$
is the averaged fraction of the $\pi^-$ momentum carried by the
gluon. For the $\omega K$ amplitudes, the term $a_3+a_5$, which is
originally negligible, is replaced by $a_3+a_5+(c_4-c_6)f_3$ and
the latter gives the significantly constructive contribution to
the rates. It can thus help understanding the reason for the
similarity between $K\omega$ and $\pi^-\omega$. On the other hand,
the subleading corrections can contribute significantly to the
processes with $\eta^{(\prime)}$ in the final states, for which
the term $a_3-a_5$ always appears in the decay amplitudes and
becomes $a_3-a_5+(c_4+c_6)f_3$ after taking into account the
corrections. We also get $a_2^{\rm SL}(J/\psi K)\approx
0.27+0.05i$ which is well consistent with the data. The result
resolves the long-standing sign ambiguity of ${\rm Re}(a_2)$.

Let us study the subleading corrections originating from the
3-parton Fock states of final-state mesons. Taking the $\omega
\pi^-$ mode as an illustration, there are two different types of
diagrams shown in Fig.~\ref{fig:3p}. In the following calculation,
we adopt the conventions $D_\alpha=\partial_\alpha +ig_s T^a
A^a_\alpha$, $\widetilde{G}_{\alpha\beta}=(1/2)
\epsilon_{\alpha\beta\mu\nu}G^{\mu\nu}$, $\epsilon^{0123}=-1$, and
use the Fock-Schwinger gauge to ensure the gauge-invariant nature
of the results
\begin{eqnarray}
A_{\mu}(x)=-\int_0^1 d v\ v G_{\mu\nu}(v x)x^\nu.
\label{eq:FSgauge}
\end{eqnarray}
For Fig.~\ref{fig:3p}(a) where the contributions come from the
3-parton Fock states of the $\omega$, because of the $V-A$
structure of the weak interaction vertex, the relevant 3-parton
light-cone distribution amplitudes (LCDA) up to the twist-4 level
are given by~\cite{Ball:1998fj}
\begin{eqnarray}
 &&\langle\omega(p_\omega,\lambda) |\bar{u}(0)\gamma_\mu
g_sG_{\alpha\beta}(vx)u(0)|0\rangle \nonumber\\
&&\cong i\frac{ f_\omega m_\omega^2}{\sqrt{2}} \int{\cal
D}\alpha\, e^{ip_\omega xv\alpha_g} \biggr\{
 \left(p^\omega_\beta g_{\alpha\mu} - p^\omega_\alpha
g_{\beta\mu}\right) \Phi(\alpha_i)  \nonumber
\\
&&  + \frac{1}{(p_\omega x)}p^\omega_\mu (p^\omega_\beta x_\alpha
-p^\omega_\alpha x_\beta  )
(\Psi(\alpha_i)-\Phi(\alpha_i))\biggr\} ,\hskip0.2cm{}
\label{eq:3pomega}
\end{eqnarray}
where ${\cal D}\alpha= d\alpha_{\bar u} d\alpha_u d\alpha_g
\delta(1-\alpha_{\bar u}-\alpha_u -\alpha_g)$, with $\alpha_{\bar
u},\alpha_u,\alpha_g$ being the fractions of the $\omega$ momentum
carried by the $\bar u$-quark, $u$-quark and gluon, respectively.
Here $\Phi$ and $\Psi$ are the twist-4 LCDAs. Note that all the
components of the coordinate $x$ should be taken into account in
the calculation before the collinear approximation is applied. The
exponential in Eq.~(\ref{eq:3pomega}) before the collinear
approximation is actually $e^{ik_g \cdot xv}$, where $k_g$ is the
$gluon$'s momentum, and the resultant calculation can be easily
performed in the momentum space with substituting $x_\alpha \to
-(i/v)(\partial/\partial k_g^\alpha)$. The result of
Fig.~\ref{fig:3p}(a) is found to be
\begin{eqnarray}
&& \langle \omega \pi^-|O_1|B^-\rangle_{\rm Fig.1(a)}
=f_{\omega}\frac{4\sqrt{2}m_\omega^2}{3 m_B^2} \nonumber\\
&&\times\langle \pi^-|\bar d \not\! p_\omega
(1-\gamma_5)b|B^-\rangle \int {\cal D}\alpha
\frac{2\Phi(\alpha_i)-\Psi(\alpha_i)}{\alpha_g}.\label{eq:3pfig1}
\end{eqnarray}
Due to G-parity, $\Phi$ and $\Psi$ are antisymmetric in
interchanging $\alpha_{\bar u}$ and $\alpha_u$ for the $\omega$,
so that Eq.~(\ref{eq:3pfig1}) vanishes.

In Fig.~\ref{fig:3p}(b), we consider the emitted gluon which
becomes a parton of the pion. We first take $G_{\mu\nu}(v x)\simeq
G_{\mu\nu}(0)e^{iv k_g^\pi \cdot x}$ and then adopt the collinear
approximation $k_g^\pi=\overline\alpha_g p_\pi$ in the final stage
of the calculation, where $\overline\alpha_g$ is the averaged
fraction of the pion's momentum carried by the gluon. The
calculation is straightforward and leads to
\begin{eqnarray}
 &&\langle \omega \pi^-|O_1|B^-\rangle_{\rm Fig.1(b)} \nonumber\\
 &&= \frac{f_\omega m_\omega}{4\sqrt{2}N_c}
 \int_0^1 dv \int_0^1 \phi_\omega(u) \nonumber\\
 &&\ \  \times
 \langle \pi^-|\bar d \gamma_\mu (1-\gamma_5) g_s \widetilde{G}_{\nu\beta} b|B^-\rangle
\frac{i\partial}{\partial{k_g^\pi}_\beta} \bigg\{ {\rm Tr} \bigg[
\not\! \epsilon^*_\omega
\nonumber\\
&&\ \ \times \bigg(
 \frac{\gamma^\nu (v \not\! k_g^\pi + u \not\! p_\omega) \gamma^\mu}
 {(v k_g^\pi +up_\omega)^2}
 - \frac{\gamma^\mu (v \not\! k_g^\pi + \bar u \not\! p_\omega) \gamma^\nu}
 {(v k_g^\pi + \bar up_\omega)^2} \bigg) \bigg] \bigg\}\nonumber\\
&& \cong -\frac{2\sqrt{2}f_{\omega}}{\overline\alpha_g
m_B^2}p_{\omega}^\alpha  \langle \pi^-|\bar d \gamma^\mu \gamma_5
g_s \widetilde{G}_{\alpha\mu} b|B^-\rangle,\label{eq:3pfig2}
\end{eqnarray}
where the $\omega$ mesons's asymptotic leading-twist distribution
amplitude $\phi_\omega(u)=6u \bar u$ has been taken and $\bar
u=1-u$. We have two unknown parameters $p_{\omega}^\alpha \langle
\pi^-|\bar d \gamma^\mu \gamma_5 g_s \widetilde{G}_{\alpha\mu}
b|B^-\rangle$ and $\overline\alpha_g$ needed to be determined.
First, let us evaluate $p_{\omega}^\alpha \langle \pi^-|\bar d
\gamma^\mu \gamma_5 g_s \widetilde{G}_{\alpha\mu} b|B^-\rangle$.
The matrix element can be calculated by considering the
correlation function:
\begin{eqnarray} &&\Pi_\alpha
(p,p+q)= i\int d^4x e^{ipx} \langle \pi^-(q)|T(j_{3p}(x)\, j_B(0))|0\rangle\nonumber\\
&& = \frac{m_B^2 f_B}{m_b} \frac{1}{m_B^2-(p+q)^2} \nonumber
\\ && \times \langle \pi^-|\bar d \gamma^\mu \gamma_5 g_s
\widetilde{G}_{\alpha\mu} b|B^-(q+p)\rangle + \cdots,
\label{eq:correlation}
\end{eqnarray}
where $j_{3p}=\bar d g_s
\widetilde{G}_{\alpha\mu}\gamma^\mu\gamma_5 b, j_B= \bar b
i\gamma_5 u$, the ellipses denote contributions from the higher
resonance states, which can couple to the current $j_B$, and the
transition matrix element can be parametrized as
\begin{eqnarray}
&& \langle \pi^-(q)|\bar d \gamma^\mu \gamma_5 g_s
\widetilde{G}_{\alpha\mu} b|B^-(p+q)\rangle \nonumber\\
&& = p_\alpha f_-(p^2) +(p_\alpha+2q_\alpha) f_+(p^2).
\end{eqnarray}
\begin{figure}
\centerline{
        {\epsfxsize1.1in \epsffile{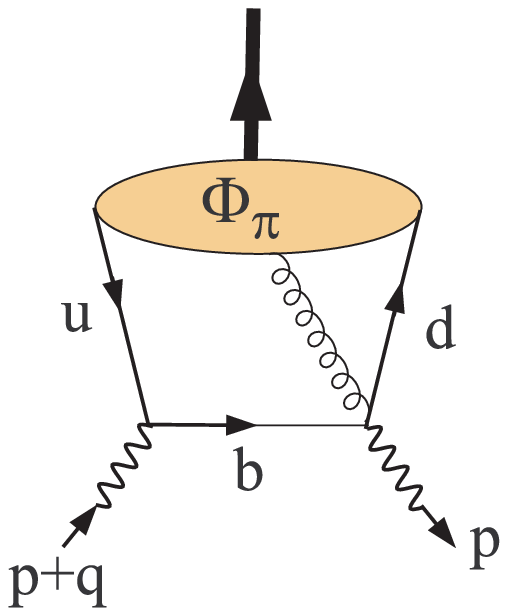}}}
\caption{The diagrammatic illustration to the correlation
function, Eq.~(\ref{eq:correlation}).}\label{fig:formfactor}
\end{figure}
In the deep Euclidean region of $(p+q)^2$, as depicted in
Fig.~\ref{fig:formfactor} the correlation function can be
perturbatively calculated in QCD and expressed in terms of
3-parton LCDAs of the pion,
 \begin{eqnarray} \Pi_{\alpha}^{\rm QCD} &=& q_{\alpha} \int_{0}^{1} \frac{du }{
m_{b}^{2} - ( p + uq)^{2}} \int_{0}^{u} d \alpha_{g} \nonumber\\
& \times & [ - 2 (p\cdot q) f_{3 \pi}  \phi_{3 \pi}
 + f_{\pi} m_{b} ( \tilde{\phi}_{\parallel} - 2 \tilde{\phi}_{\perp})
 ],
\label{eq:Tqcd}\end{eqnarray}
where $u=\alpha_d+\alpha_g$, and the
3-parton pion LCDAs are defined
by~\cite{Braun:1989iv,Belyaev:1994zk}
\begin{eqnarray}
\lefteqn{ \langle\pi(q) |\bar{d}(x) g_s
G_{\mu\nu}(vx)\sigma_{\alpha\beta}\gamma_5u(0)|0\rangle }
\nonumber\\
&=&if_{3\pi}[q_\beta (q_\mu g_{\nu\alpha}-q_\nu g_{\mu\alpha}) -
q_\alpha(q_\mu g_{\nu\beta}-q_\nu g_{\mu\beta})]
\nonumber\\
&& \times \int{\cal
D}\alpha\,\phi_{3\pi}e^{iqx(\alpha_d+v\alpha_g)}\,,
\label{3pidefinition} \end{eqnarray}
\begin{eqnarray} && \langle\pi(q) |\bar{d}(x)\gamma_\mu
g_s\tilde{G}_{\alpha\beta}(vx)u(0)|0\rangle= if_\pi\left( q_\alpha
g_{\beta\mu} - q_\beta
g_{\alpha\mu}\right) \nonumber\\
&& \times \int{\cal D}\alpha\,\tilde{\phi}_\perp
e^{iqx(\alpha_d+v\alpha_g)} - if_\pi\frac{q_\mu}{qx}(q_\alpha
x_\beta -q_\beta x_\alpha ) \nonumber\\
&& \times \int{\cal
D}\alpha\,(\tilde{\phi}_\parallel+\tilde{\phi}_\perp)
e^{iqx(\alpha_d+v\alpha_g)}\,. \label{qqgpi}
\end{eqnarray}
Here $\phi_{3\pi}$ is a twist-3 DA, $\tilde{\phi}_\perp$ and
$\tilde{\phi}_\parallel$ are all of twist-4,
\begin{eqnarray}
\phi_{3\pi}(\alpha_i)&=&360 \alpha_d\alpha_{\bar u}\alpha_g^{2}
\Big[1+\omega_{1,0}\frac12(7\alpha_g-3) \nonumber
\\
&&{}+\omega_{2,0}(2-4\alpha_d\alpha_{\bar
u}-8\alpha_g+8\alpha_g^{2}) \nonumber\\
&&{}+\omega_{1,1}(3\alpha_d \alpha_{\bar
u}-2\alpha_g+3\alpha_g^{2})\Big]
~,\nonumber\\
\tilde{\phi}_\perp (\alpha_i)&=&
30\delta^2\alpha_g^{2}(1-\alpha_g)[\frac13+2 \varepsilon
(1-2\alpha_g)] ~, \nonumber
\\
\tilde{\phi}_\parallel
(\alpha_i)&=&-120\delta^2\alpha_d\alpha_{\bar u}\alpha_g[\frac13+
\varepsilon (1-3\alpha_g)] ~. \label{3pi}
\end{eqnarray}
Since the quark's momentum after emitting the gluon is roughly of
order $m_B/2$ and the emitted gluon's momentum is
$\Lambda_\chi\sim \overline \alpha_g p_\pi$ ($\overline\alpha_g$
will be discussed below), we set the scale for the separation of
the perturbative and nonperturbative parts at $\mu_h=
\sqrt{\Lambda_\chi m_B/2}\simeq 1.4$~GeV.   The corresponding
parameters at the scale $\mu_h$ read: $f_{3\pi}=0.0032$~GeV$^2$,
$\omega_{1,0}=-2.63$, $\omega_{2,0}=9.62$, $\omega_{1,1}=-1.05$,
$\delta^2=0.19$ GeV$^2$, $\varepsilon
=0.45$~\cite{Belyaev:1994zk}.
 To calculate $f_\pm$, the contributions of
higher resonances in Eq.~(\ref{eq:correlation}) are approximated
by
 \begin{equation}
\frac{1}{\pi}\int^\infty_{s_0} \frac{\im \Pi_{\alpha}^{\rm
QCD}}{s-(p+q)^2}ds\,,
 \end{equation}
 where $s_0$ is the threshold of higher resonances.
Equating Eqs.~(\ref{eq:correlation}) and (\ref{eq:Tqcd}) and
making the Borel transformation: ${\cal B}[m_{B}^{2} -
(p+q)^{2}]^{-1}=\exp(- m_{B}^{2}/M^{2})$, we obtain the light-cone
sum rule
\begin{eqnarray} \lefteqn{f_{-}(p^{2}) =
\frac{m_{b}}{2 m_{B}^{2} f_{B}} \int_0^{1} du \int_{\Delta}^{u} d
\alpha_{g} \; e^{( \frac{m_{B}^{2}}{M^{2}} - \frac{m_{b}^{2} -
\bar{u} p^{2}}{u M^{2}}) } }\nonumber\\
&&\times \left[ f_{3 \pi} \frac{m_{b}^{2} - p^{2}}{u^{2}} \phi_{ 3
\pi} - f_{\pi} \frac{m_{b}}{u} ( \tilde{\phi}_{\parallel} - 2
\tilde{\phi}_{\perp}) \right],
\end{eqnarray}
and $f_{+}(p^{2})=- f_{-}(p^{2})$, where $\Delta=u - (m_{b}^{2} -
p^{2})/( s_{0} - p^{2})$.   Using the above parameters for LCDAs,
$m_{b} = (4.7\pm 0.1)$~GeV, and $f_{B} = 180$~MeV, we obtain the
stable $f_\pm$ prediction by adopting $ s_{0} \simeq 37$~GeV$^2$
and $M^2 \approx (9-20)$~GeV$^2$. The result is depicted in
Fig.~\ref{fig:lcsr}(a). The resulting value is $f_\mp
(m_\omega^2)= \pm(0.057\pm 0.005)~{\rm GeV^2}$, where the
uncertainty comes from the sum rule analysis.  We then get
\begin{eqnarray}
&& p_{\omega}^\alpha \langle \pi^-|\bar d \gamma^\mu \gamma_5 g_s
\widetilde{G}_{\alpha\mu} b|B^-\rangle \nonumber\\
&& = -f_-(m_\omega^2)\times (m_B^2-m_\pi^2-m_\omega^2) \simeq -1.6
{\rm\ GeV}^4. \hskip0.5cm{}
\end{eqnarray}

Next, we determine the value of $\overline \alpha_g$. For
illustration, we plot in Fig.~\ref{fig:lcsr}(b) the amplitude
$A_{f_-}$ of the $f_-(m_\omega^2)$ sum rule versus $\alpha_g$ and
$u(=\alpha_d+\alpha_g)$ by adopting $M^2$=10~GeV$^2$, where
$A_{f_-}$ satisfies $f_-=\int^1_0 du \int^u_0 d\alpha_g A_{f_-}$,
i.e. the volume in the plot is equal to $f_-(m_\omega^2)$. The
resultant form factor is dominated by the region where $u\gtrsim
60\%, \alpha_g\lesssim 30\%$. The averaged fraction of the pion
momentum carried by the gluon is then estimated to be $\overline
\alpha_g=(\int^1_0 du \int^u_0 d\alpha_g \alpha_g
A_{f_-})/f_-\simeq$~0.23.

We therefore obtain $\langle \omega \pi^-|O_1|B^-\rangle_{\rm
Fig.1(b)}\simeq 0.13$ and $f_3=0.12$ which gives the correction to
$a_i$ as defined in Eq.~(\ref{eq:wc}). We list $a_i$ without and
with the subleading corrections in Table~\ref{tab:wcai}, where the
approximation $-\sqrt{2}i p_{\pi}^\alpha \langle \omega|\bar u
\gamma^\mu g_s \widetilde{G}_{\alpha\mu} b|B^-\rangle /
A_0^{B\omega} \simeq p_{\omega}^\alpha \langle \pi^-|\bar d
\gamma^\mu \gamma_5 g_s \widetilde{G}_{\alpha\mu} b|B^-\rangle /
F_1^{B\pi}$ has been made. Note that $a_6$ and $a_8$ do not
receive subleading corrections.

\begin{figure}[t!]
\centerline{\epsfig{file=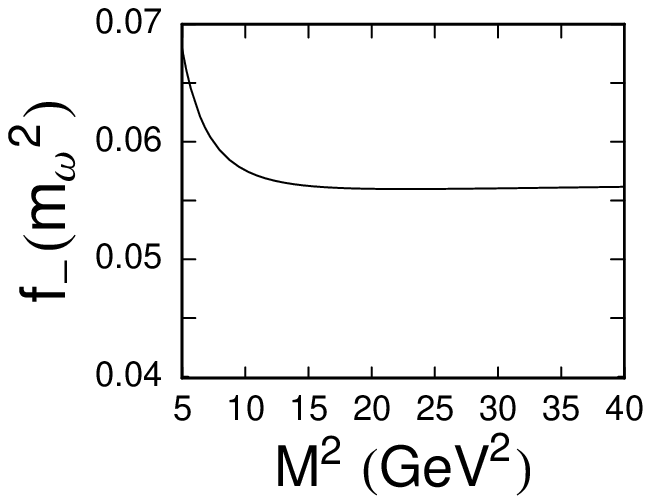,clip=46mm, height=30mm}\hskip1mm
\epsfig{file=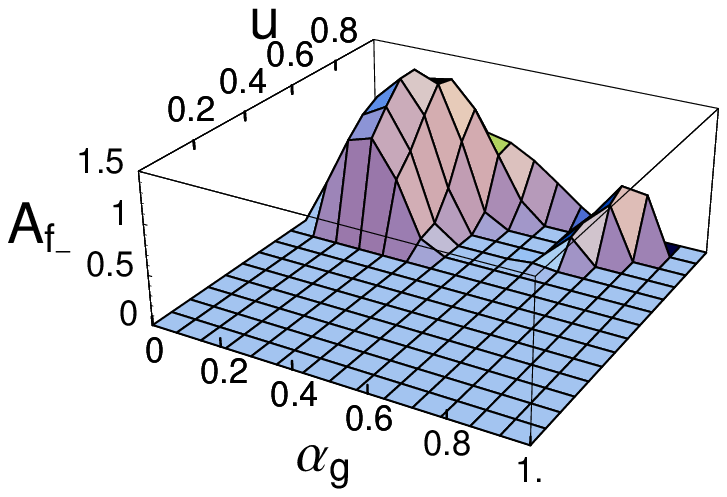,clip=46mm, height=30mm}} \caption{(a)
Form factor $f_-(m_\omega^2)$ plotted as a function of the Borel
mass squared $M^2$. (b) $f_-(m_\omega^2)=\int^1_0 du \int^u_0
d\alpha_g A_{f_-}$ with $M^2$=10 GeV$^2$. The volume in the plot
is equal to $f_-(m_\omega^2)$. Here $u=\alpha_d+\alpha_g$.}
\label{fig:lcsr}
\end{figure}

{\squeezetable
\begin{table*}[t]
\vspace{0.3truecm}
\begin{center}
\caption{Values for $a_i$ for charmless $B$ decay processes
without (first row) and with (second row) 3-parton Fock state
contributions of final state mesons, where $a_{3{\rm -}10}$ are in
units of $10^{-4}$ and the annihilation effects are not
included.}\label{tab:wcai}
\begin{tabular}{ c c c c c c c c c c}
\hline \hline
 $a_1$  & $a_2$ & $a_3$ & $a_4$ & $a_5$  & $a_6$ & $a_7$ & $a_8$ &
 $a_9$  & $a_{10}$  \\ \hline
 $1.02+0.014i$   & $0.10-0.08i$ & $26+26i$  & $-328 - 91i$ & $1.2-30i$  & $-487 - 72i$ &
 $0.7 + 0.3i$ & $4.5 + 0.6i$ & $-89 - 0.1i$ & $-5.9+ 7i$ \\
 $0.974+0.014i$   & $0.25-0.08i$ & $-55+26i$  & $-291 - 91i$ & $112-30i$  & $-487 - 72i$ &
 $-0.3+ 0.3i$ & $4.5 + 0.6i$ & $-88 - 0.1i$ & $-18+ 7i$ \\
 \hline\hline
\end{tabular}
\end{center}
\end{table*}}

\begin{figure}[tb]
\centerline{
{\epsfxsize40mm \epsffile{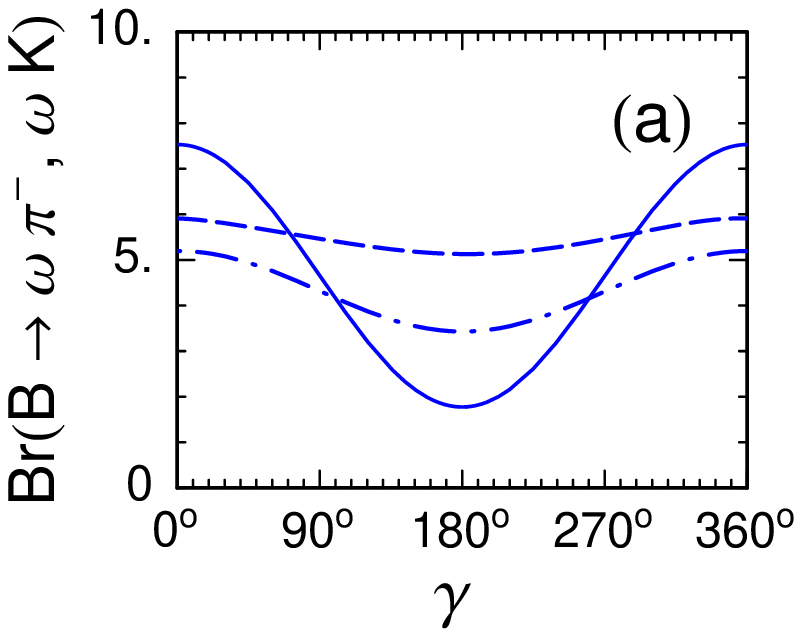}} {\epsfxsize40mm
\epsffile{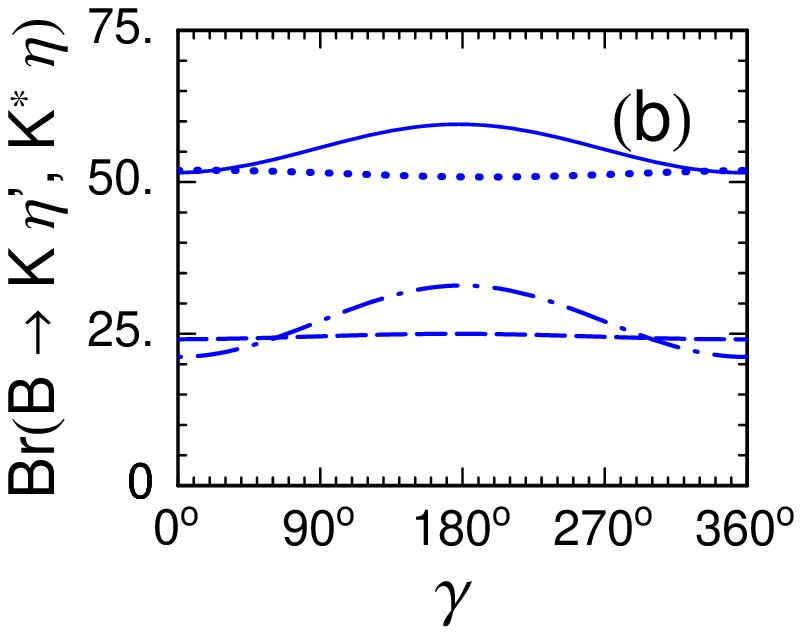}}} \caption{(a) Dash, solid, and dot-dash for
$\overline B\to \omega\pi^-, \omega K^-$ and $\omega \overline
K^0$; (b) solid, dots, dot-dash, and dash for $K^-\eta',\overline
K^0\eta', K^{*-}\eta$, and $\overline K^{*0}\eta$. Brs are in
units of $10^{-6}$.} \label{fig:omegakpi}
\end{figure}

In the following analysis, the LC sum rule form factors and
$m_s=80$~MeV are used. We will instead use a smaller
$A_0^{B\rho}=0.28$ which is preferred by the $\omega \pi^-$ data.
The $\omega \pi^-$ mode is ideal for extracting $A_0^{B\rho}$
since its rate is insensitive to annihilation effects. We find
that the spectator parameter $X_{H}= \ln
\frac{m_B}{\Lambda_{h}}(1+\rho_{H}\,e^{i\phi_{H}})$ is consistent
with zero in the analysis. The reason is that since the spectator
interaction with a gluon exchange between the emitted meson and
the recoiled pseudoscalar meson of twist-3 LCDA $\Phi_\sigma$ is
end-point divergent in the collinear expansion, the vertex of the
gluon and spectator quark should be considered inside the pion
wave function, i.e., for this situation the pion itself is at a
3-parton Fock state. Annihilation effects have been emphasized in
$\phi K$ studies~\cite{Cheng:2000hv}. We adopt the annihilation
parameters $\rho_A\simeq 0.9, \phi_A\simeq 0$ which give
$Br(B^-\to \phi K^-)\simeq 8.5\times 10^{-6}$, consistent with the
current data. Here the annihilation parameter of $VP$ modes is
defined as $X_{A}^{VP}= \ln \frac{m_B}{\Lambda_{h}}
(1+\rho_{A}\,e^{i\phi_{A}})$~\cite{Beneke:2001ev} and its the
imaginary part is neglected since the BRs are insensitive to it.
In Fig.~\ref{fig:omegakpi}(a) we plot the BRs of $\omega \pi^-$
and  $\omega K$ modes versus $\gamma$ ($\equiv {\rm
arg}V^*_{ub}$). The results for $\gamma\approx (60-120)^\circ$ are
in good agreement with data. At $\gamma=90^\circ$, it gives
$\omega \pi^-, \omega K^-,\omega \overline K^0$ to be $5.5, 4.5,
4.3$, respectively, in units of $10^{-6}$. Without the
contributions from 3-parton Fock states of mesons, $\phi K^-,
\omega \pi^-, \omega K^-,\omega \overline K^0$ will become $11,
3.9, 3.1, 2.9$~(in units of 10$^{-6})$. The 3-parton Fock state
effects give constructive contributions to $\omega \pi, \omega K$
modes, but destructive one to the $\phi K$ mode.

\begin{figure}[tb]
\centerline{ \epsfig{file=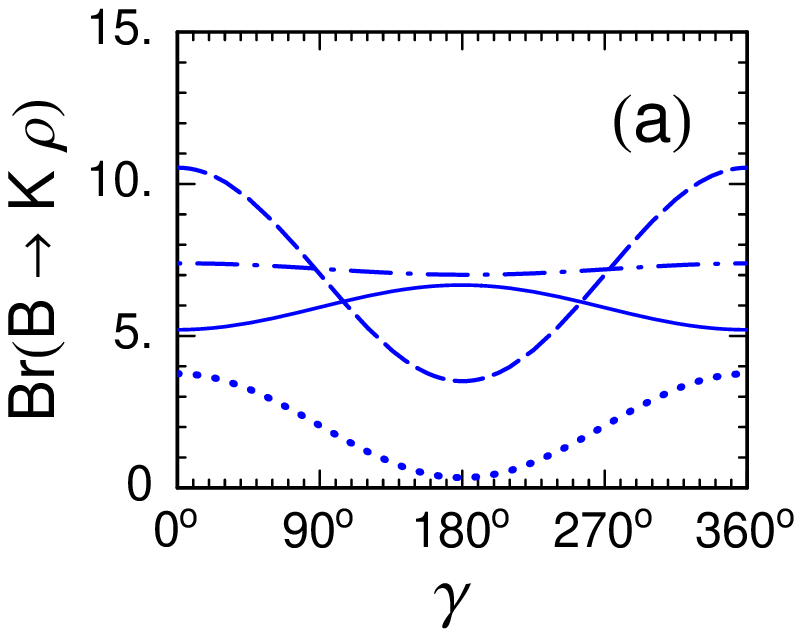,width=40mm}
\epsfig{file=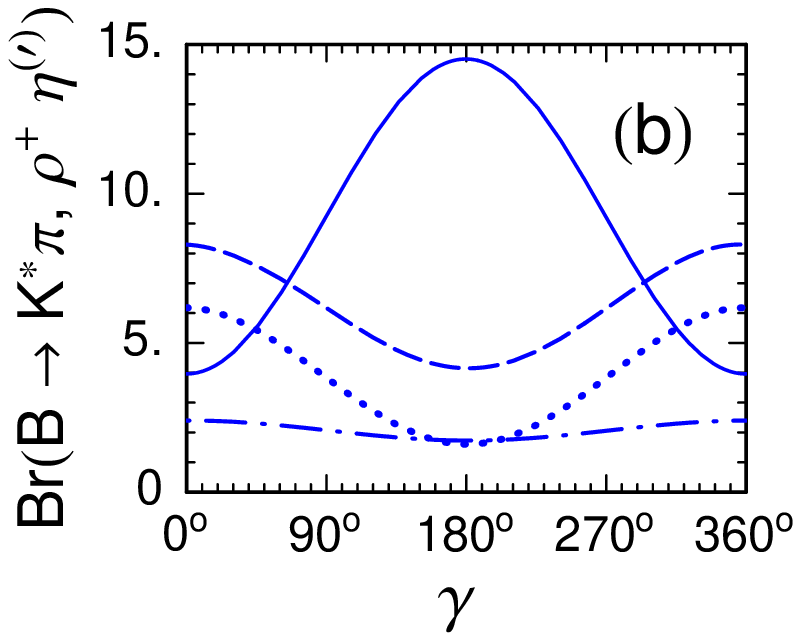,width=40mm} } \caption{(a) Solid, dash,
dot-dash, and dots for $\overline B\to \overline K^0\rho^0,
K^-\rho^+, \overline K^0\rho^-$, and $K^-\rho^0$; (b) solid,
dot-dash, dash, and dots for $\overline B\to K^{*-}\pi^+,
\overline K^{*0}\pi^0, \rho^-\eta$ and $\rho^-\eta'$. Brs are in
units of $10^{-6}$. } \label{fig:vp}
\end{figure}

The corrections from 3-parton Fock states of the kaon also give a
definite answer to the longstanding problem for $a_2(J/\psi K)$.
In the earlier study, to account for the experimental value
$|a_2|$, the parameter $\rho_H$ has to be $\gtrsim
1.5$~\cite{Cheng:2000kt}. As emphasized in passing, without
fine-tuning $\rho_H$, we calculate the amplitudes from the
3-parton Fock states of the kaon. With the same procedure as shown
above, we obtain $f_\mp (m_{J/\psi}^2)\simeq \pm 0.08,\, f_3\simeq
0.14$ and thus $a_2^{\rm SL}=a^{t2}_2+c_1(\mu_h) f_3\simeq
0.10+0.05i+ c_1(\mu_h) 0.14= 0.27+ 0.05i$,  where $a^{t2}_2$ is
determined up to the twist-2 order and the SU(3) approximation for
$f_3$ has been made. The result for $a_2^{\rm SL}$  is well
consistent with that extracted from data. This solves the
long-standing sign ambiguity of $a_2(J/\psi K)$ which turns out to
be positive for its real part. Note that if ${\rm Re}(a_2)$ were
negative, $f_3$ would have to be $\sim -0.3$ which in turns would
lead to $\phi K\sim 20\times 10^{-6}$ and $\omega \pi^-, \omega K
\sim 1\times 10^{-6}$!

The subleading corrections could give significant contributions to
the decays with $\eta^{(\prime)}$ in the final states because
these decay amplitudes always contain the singlet factor
$a_3-a_5$. We plot the BRs of $K\eta^{\prime}, K^*\eta$ modes
versus $\gamma$ in Fig.~\ref{fig:omegakpi}(b), where
$X_A^{PP}\approx 0$, the annihilation parameter for $PP$ modes,
has been used as it could give good fit results for $K\pi, \pi\pi$
rates~\cite{Beneke:2002nj}. We do not consider the singlet
annihilation correction~\cite{Beneke:2002jn} in $K\eta'$ modes
because it is still hard to determine at present.  With (without)
the subleading corrections, we see that $K^-\eta' \gtrsim
\overline K^0\eta' \approx 55\, (35)$, and $K^{*-}\eta \gtrsim
\overline K^{*0}\eta \approx 24\, (20)$, in units of $10^{-6}$.
The corrections give 70\% and 25\% enhancements to $K \eta'$ and
$K^* \eta$ rates, respectively. Note that with (without) the
corrections, $K^{*}\eta' \lesssim 1\times 10^{-6}$ $(\gtrsim
4\times 10^{-6})$ and $K\eta \lesssim 1\times 10^{-6}$ $(\gtrsim
1\times 10^{-6})$.
In pQCD calculation~\cite{Kou:2001pm}, it seems that $\overline
K^0 \eta^\prime(= 41\times 10^{-6})$ was underestimated while
$\overline K^0 \eta(=7\times 10^{-6})$ overestimated compared to
the data\cite{belle,babar_omegaK}. Within the QCDF framework, by
only considering the two-parton LCDAs of mesons, Beneke and
Neubert~\cite{Beneke:2002jn} obtained $K^* \eta \sim 13\times
10^{-6}$, just half of the experimental value, but with a huge
error.

 For further comparison with other calculations, we
plot $K\rho, K^*\pi, \rho^-\eta^{(\prime)}$ modes in
Fig.~\ref{fig:vp}. The subleading contributions to these BRs are
$\lesssim 15\%$. At $\gamma=90^\circ$, we have $\overline
K^0\rho^0, K^-\rho^+, \overline K^0 \rho^-, K^-\rho^0
=6,7,7,2~(\times 10^{-6})$, and $K^{*-}\pi^+, \overline K^{*0}
\pi^0, \rho^-\eta, \rho^-\eta'=9,2,6,4~(\times 10^{-6})$. In
\cite{Du:2002up}, to fit $\phi K,\omega K$ rates, the annihilation
effects dominate the decay amplitudes and the form factors
$F^{BK}_1, A_0^{B\pi}$ are rather small, such that it will lead to
$\rho^+ K^{-,0}: \rho^0 K^{-,0}: \omega K^{-,0}\approx
1:(1/\sqrt{2})^2:(1/\sqrt{2})^2$, while the pQCD results for rates
are $\overline K^0\rho^0, K^-\rho^+,\overline K^0 \rho^-,K^-\rho^0
=2.5,5.4,3.0,2.2\, (\times 10^{-6})$~\cite{Chen:2001jx}.

In conclusion, we have calculated the  contributions arising from
3-parton Fock states of mesons in $B$ decays. We find that the
contributions could give significant corrections to decays with an
$\omega$ or $\eta^{(\prime)}$ in the final states. Our main
results for $\gamma\approx (60-110)^\circ$ are $\omega \pi^-,
\omega K^-,\omega \overline K^0 \approx 6.0, (6\sim 5), 5.1$,
respectively, in units of $10^{-6}$, while the previous QCDF
global fit to $VP$ modes and the pQCD results gave smaller $\omega
K$ BRs of order $\lesssim 3\times
10^{-6}$~\cite{Aleksan:2003qi,Chen:2001jx}. We predict that
$\overline K^0\rho^0\sim \omega K$ and $\overline
K^0\rho^0/K^-\rho^0\approx 3$. Including the corrections we obtain
$a_2(J/\psi K)\approx 0.27+0.05i$ which is well consistent with
the data. The sign of ${\rm Re}(a_2)$ turns out to be positive.
Without resorting to the unknown singlet annihilation effects,
3-parton Fock state contributions can enhance $K\eta'$ to the
level above $5\times 10^{-5}$.

  This work was supported in part by the grant from
NSC91-2112-M-033-013. I am grateful to H.Y. Cheng for a critical
reading of the manuscript and useful comments.


\begin{thebibliography}{99}
\bibitem{Beneke:2001ev}
M.~Beneke {\it et al.},
Nucl.\ Phys.\ B {\bf 606}, 245 (2001).

\bibitem{Beneke:2002nj}
M.~Beneke,
arXiv:hep-ph/0207228.

\bibitem{Keum:2000ph}
Y.Y.~Keum, H.n.~Li and A.I.~Sanda,
Phys.\ Lett.\ B {\bf 504}, 6 (2001).

\bibitem{Cheng:2000hv}
H.Y.~Cheng and K.C.~Yang,
Phys.\ Rev.\ D {\bf 64}, 074004 (2001).

\bibitem{belle} Belle Collaboration, R.S.~Lu {\it et al.},
Phys.\ Rev.\ Lett.\  {\bf 89}, 191801 (2002);
Belle Collaboration, H. Aihara,
talk presented at Flavor Physics and CP Violation, Paris,
France, June, 3-6, 2003.

\bibitem{babar_omegaK} BaBar Collaboration, M. Pivk, talk presented
at the XXXVIII  Rencontres de Moriond, Les Arcs, Savoie, France,
March 15-29, 2003.


\bibitem{Du:2002up}
D.S.~Du {\it et al.},
Phys.\ Rev.\ D {\bf 65}, 094025 (2002).


\bibitem{Du:2002cf}
D.S.~Du {\it et al.},
Phys.\ Rev.\ D {\bf 67}, 014023 (2003).

\bibitem{Chen:2001jx}
C.H.~Chen,
Phys.\ Lett.\ B {\bf 525}, 56 (2002).

\bibitem{Aleksan:2003qi}
R.~Aleksan {\it et al.},
Phys.\ Rev.\ D {\bf 67}, 094019 (2003). 



\bibitem{Ball:1998fj}
P.~Ball and V.M.~Braun,
arXiv:hep-ph/9808229.

\bibitem{Braun:1989iv}
V.M.~Braun and I.B.~Filyanov,
Z.\ Phys.\ C {\bf 48}, 239 (1990).

\bibitem{Belyaev:1994zk}
V.M.~Belyaev {\it et al.},
Phys.\ Rev.\ D {\bf 51}, 6177 (1995).


\bibitem{Cheng:2000kt}
H.Y.~Cheng and K.C.~Yang,
Phys.\ Rev.\ D {\bf 63}, 074011 (2001).

\bibitem{Beneke:2002jn}
M.~Beneke and M.~Neubert,
Nucl.\ Phys.\ B {\bf 651}, 225 (2003). 

\bibitem{Kou:2001pm}
E.~Kou and A.~I.~Sanda,
Phys.\ Lett.\ B {\bf 525}, 240 (2002). 

\end{thebibliography}
\end{document}